\documentstyle[12pt]{article}

\font\Bbbfont=msbm10
\def\Bbb#1{\hbox{\Bbbfont#1}}

\begin{document}

\hfill DFNT--T  9/99

\hfill SISSA 48/99/FM

\begin{center}
{ \Large \bf  Invariants of spin networks 
with boundary\\
in Quantum Gravity and TQFT's}

\vspace{24pt}

{\large
{\sl Gaspare Carbone}$\,^{a}$,
{\sl Mauro Carfora}$\,^{b}$
and
{\sl \underline {Annalisa Marzuoli}}$\,^{b}$}
\vspace{24pt}

$^a$~S.I.S.S.A.-I.S.A.S.,\\
Via Beirut 2-4, 34013 Trieste, Italy

\vspace{12pt}
$^b$~Dipartimento di Fisica Nucleare e Teorica, \\
Universit\`a degli Studi di Pavia, \\
via A. Bassi 6, I-27100 Pavia, Italy, \\
and\\
Istituto Nazionale di Fisica Nucleare, Sezione di Pavia, \\
via A. Bassi 6, I-27100 Pavia, Italy

\end{center}

\vspace{12pt}

\begin{center}
{\bf Abstract}
\end{center}

\vspace{12pt}

The search for classical or quantum combinatorial invariants of compact $n$-dimensional manifolds 
$(n=3,4)$ plays a key role both in
topological field theories and in lattice quantum gravity (see {\em e.g.} \cite{ponzano}, \cite{turaev}, \cite{ooguri4d},
\cite{carter98}).
We present here a generalization of the partition function proposed by Ponzano and Regge  to the case of
a compact $3$-dimensional simplicial pair $(M^3, \partial M^3)$. The resulting state sum $Z[(M^3, \partial M^3)]$
contains both Racah--Wigner $6j$ symbols associated with tetrahedra and Wigner $3jm$ symbols associated with 
triangular faces lying in $\partial M^3$.
The analysis of the algebraic identities associated with the combinatorial transformations involved in the proof
of the topological invariance makes it manifest a common structure underlying the $3$-dimensional models
with empty and non empty boundaries respectively. The techniques developed in the $3$-dimensional case
can be further extended in order to deal with combinatorial models in $n=2,4$ and possibly to establish a hierarchy
among such models. As an example we derive here a $2$-dimensional closed state sum model including suitable sums of
products of double $3jm$ symbols, each one of them being associated with a triangle in the surface.

\vskip 1 cm

Recall that a closed $PL$-manifold of dimension $n$ is a polyhedron $M \cong |T|$, each point of which has a neighborhood, in $M$,
$PL$-homeomorphic to an open set in ${\Bbb R}^n$.
The symbol $\cong$ denotes homeomorphism, $T$ is the underlying (finite) simplicial complex and $|T|$ denotes the associated 
topological space, namely the set theoretic union of all simplices of $T$ endowed with its natural topology.\par
\noindent $PL$-manifolds are realized by simplicial manifolds under the equivalence relation generated by 
$PL$-homeomorphisms. In particular,
two $n$--dimensional closed $PL$-manifolds $M_1 \cong |T_1|$ and $M_2 \cong |T_2|$ are $PL$-homeomorphic, or
$M_1 \:\cong_{PL} \:M_2$, 
if there exists a map $g:M_1 \rightarrow M_2$ which is both a homeomorphism and a simplicial
isomorphism (see {\em e.g.} \cite{rourke} for more details on both this definition and other issues from $PL$--topology
which will be used in the following).\par
\noindent We shall use the notation\par

\begin{equation}
T \:\longrightarrow \: M\:\cong \:|T|
\label{clotri}
\end{equation}

\noindent to denote a {\em particular triangulation} of the closed $n$-dimensional $PL$-manifold $M$.\par 
\noindent In order to extend the previous notation
to the case of a  $PL$-pair $(M,\partial M)$ of dimension $n$, recall that a simplicial complex is {\em pure} provided 
that all its facets (namely its faces of maximal dimension) have the same dimension. Moreover,
the {\em boundary complex} of a pure simplicial $n$-complex $T$ is denoted by $\partial T$ and it is the
subcomplex of $T$ the facets of which are the $(n-1)$-faces of $T$ which are contained in only one facet of $T$.
The set of the interior faces of $T$ is denoted by $int(T) \doteq T \setminus \partial T$. Then:\par

\begin{equation}
(T, \partial T )\:\longrightarrow \: (M, \partial M)\:\cong \:(|T|, |\partial T|)
\label{bountri}
\end{equation}

\noindent denotes a triangulation on $(M, \partial M)$, where $\partial T$ is the unique triangulation 
induced on the $(n-1)$-dimensional boundary $PL$-manifold $\partial M$ by the chosen
triangulation $T$ in $M$.\par

\vskip 1 cm 

Following \cite{carbone}, the connection between a recoupling scheme of $SU(2)$ angular momenta and the 
combinatorial structure of a compact, 
$3$-dimensional simplicial pair $(M^3, \partial M^3)$ can be established by considering {\em colored} triangulations which
allow to specialize the map (\ref{bountri}) according to:\par

\begin{equation}
(T^3(j), \partial T^3 (j', m))\:\longrightarrow\:(M^3, \partial M^3)
\label{coltri}
\end{equation}

\noindent This map represents a triangulation associated with an admissible assignement of both spin variables 
to the collection of the edges 
in $(T^3, \partial T^3)$ and of momentum projections to the subset of edges lying in $\partial T^3$. 
The collective variable $j \equiv \{j_A\}$, $A=1,2,\ldots,N_1$, denotes all the spin variables, $n'_1$ of which are associated
with the edges in the boundary (for each $A$: $j_A=0,1/2,1,3/2,\ldots$ in  $\hbar$ units).
Notice that the last subset is labelled both by $j' \equiv \{j'_C \}$, $C=1,2,\ldots,n'_1$,
and by $m \equiv \{m_C\}$, where $m_C$ is the projection of $j'_C$ along the fixed reference axis (of course, for each $m$,
$-j \leq m \leq j$ in integer steps). The consistency in the assignement of the $j$, $j'$, $m$ variables is ensured if we require 
that:\par 

\begin{itemize}

\item each $3$-simplex $\sigma_B^3$, ($B=1,2,\ldots,N_3$), 
in $(T^3, \partial T^3)$ must be associated, apart from a phase factor, with a $6j$ symbol of $SU(2)$, namely\par

\begin{equation}
\sigma^3_B \:\longleftrightarrow \:(-1)^{\sum_{p=1}^6 j_p} \:
\left\{ \begin{array}{ccc}
j_1 & j_2 & j_3 \\
j_4 & j_5 & j_6
\end{array}\right\}_B
\label{tresim}
\end{equation}

\item each $2$-simplex
$\sigma_D^2$, $D=1,2,\ldots,n'_2$ in $\partial T^3$ must be associated with a Wigner $3jm$ symbol of $SU(2)$ according to\par

\begin{equation}
\sigma_D^2 \:\longleftrightarrow\: (-1)^{(\sum_{s=1}^3 m_s)/2}
\left( \begin{array}{ccc}
j'_1 & j'_2 & j'_3 \\
m_1 & m_2 & -m_3
\end{array}\right)_D
\label{duesim}
\end{equation}

\end{itemize}

\noindent Then the following state sum can be defined:\par

\begin{eqnarray}
\lefteqn{Z[(M^3, \partial M^3)]\,=}\nonumber\\
& & =\,\lim_{L\rightarrow \infty}\:
\sum_
{\left \{\begin{array}{c}
(T^3(j), \partial T^3(j',m))\\ 
j,j',m \leq L
\end{array}\right\}}
Z[(T^3(j),\partial T^3(j',m)) \rightarrow (M^3, \partial M^3); L]
\label{clsum1}
\end{eqnarray} 

\noindent where:\par

\begin{eqnarray}
\lefteqn{Z[(T^3(j),\partial T^3(j',m)) \rightarrow (M^3, \partial M^3); L] =}\hspace{.5in}\nonumber \\
& & =\Lambda(L)^{-N_0}\,\prod_{A=1}^{N_1} (-1)^{2j_A} (2j_A+1)\,\prod_{B=1}^{N_3} (-1)^{\sum_{p=1}^6 j_p}
\left\{ \begin{array}{ccc}
j_1 & j_2 & j_3 \\
j_4 & j_5 & j_6
\end{array}\right\}_B \cdot  \nonumber \\
& & \cdot \,\prod_{D=1}^{n'_2} (-1)^{(\sum_{s=1}^3 m_s)/2}
\left( \begin{array}{ccc}
j'_1 & j'_2 & j'_3 \\
m_1 & m_2 & -m_3
\end{array}\right)_D 
\label{clsum2}
\end{eqnarray}

\noindent $N_0$, $N_1$, $N_3$ denote respectively the total number of vertices, edges and tetrahedra in $(T^3(j),
\partial T^3(j',m))$, while $n'_2$ is the number of $2$-simplices lying in $\partial T^3(j',m)$. Notice that there appears 
a factor $\Lambda(L)^{-1}$ for each vertex in $\partial T^3(j',m)$, with $\Lambda (L)\equiv 4L^3/3C$, $C$ an arbitrary constant.
It is worthwile to remark also that  products of $6j$ and $3jm$ 
coefficients  of the kind which appear in (\ref{clsum2}) are known as $jm$ {\em coefficients} in the quantum theory
of angular momentum (see {\it e.g} \cite{lituani}). Their semiclassical limit can be defined in a consistent way by requiring
that simultaneously $j,j' \rightarrow \infty$ and $m \rightarrow \infty$ with the constraint $-j' \leq m \leq j'$.
The summation in (\ref{clsum1}) has precisely this meaning , apart from the introduction of the cut--off $L$. \par
\noindent 
The state sum given in (\ref{clsum1}) and (\ref{clsum2}) 
when $\partial M^3 =\emptyset$ reduces to the usual Ponzano--Regge partition function for the 
closed manifold $M^3$ (see \cite{ponzano}); in such a case, it can be rewritten here as:\par

\begin{equation}
Z[M^3]\,=\,\lim_{L\rightarrow \infty}\:\sum_{\{T^3(j), j \leq L\}}\:Z[T^3(j)\rightarrow M^3; L]
\label{prsum1}
\end{equation}

\noindent where the sum is extended to all assignements of spin variables such that each of them is not greater than
the cut--off $L$, and each term under the sum is given by:\par

\begin{equation}
Z[T^3(j)\rightarrow M^3; L] =\Lambda(L)^{-N_0}\,\prod_{A=1}^{N_1}(-1)^{2j_A} (2j_A+1)\prod_{B=1}^{N_3} (-1)^{\sum_{p=1}^6 j_p}
\left\{ \begin{array}{ccc}
j_1 & j_2 & j_3 \\
j_4 & j_5 & j_6
\end{array}\right\}_B
\label{prsum2}
\end{equation}

\noindent As is well known, the above state sum gives the semiclassical partition function of Euclidean gravity with
an action discretized according to Regge's prescription \cite{regge}.\par

\vskip 1 cm

The state sum given in (\ref{prsum1})
and (\ref{prsum2}) is formally invariant under a set of topological  transformations performed 
on $3$-simplices in $T^3(j)$: following Pachner \cite{pachner87},
 they are commonly known as {\em bistellar elementary operations} or {\em bistellar moves}.
It is a classical result  (see {\it e.g.} \cite{ponzano} and \cite{carter95}) 
that such moves can be expressed algebraically in terms of the Biedenharn--Elliott identity
(representing the moves ($2$ tetrahedra) $\leftrightarrow$ ($3$ tetrahedra)) and of both the B-E identity and the 
orthogonality conditions (which represent the moves ($1$ tetrahedron) $\leftrightarrow$ ($4$ tetrahedra)). 
The expression of the Biedenharn--Elliott identity reads:\par

\begin {eqnarray}
\lefteqn{\sum_{X} (2X+1)(-1)^{\Theta +X}
\left\{ \begin{array}{ccc}
a & b & X \\
c & d & p
\end{array} \right\}
\left\{ \begin{array}{ccc}
c & d & X \\
e & f & q
\end{array} \right\}
\left\{ \begin{array}{ccc}
e & f & X \\
b & a & r
\end{array}\right\}\,=}\hspace{1.5in} \nonumber\\
& &  =\,\left\{ \begin{array}{ccc}
p & q & r \\
e & a & d
\end{array} \right\}
\left\{ \begin{array}{ccc}
p & q & r \\
f & b & c
\end{array} \right\}
\label{beidentity}
\end{eqnarray}

\noindent where now $a,b,c,\ldots$ denote angular momentum variables, $(2X+1)$ is the dimension of the irreducible representation
of $SU(2)$ labelled by $X$, and $\Theta=a+b+c+d+e+f+p+q+r$. The orthogonality conditions amount to:

\begin{equation}
\sum_X (2X+1)
\left\{ \begin{array}{ccc}
a & b & X \\
c & d & e
\end{array}\right\}\,
\left\{ \begin{array}{ccc}
a & b & X \\
c & d & f
\end{array}\right\}
\:=\:(2e+1)^{-1}\,\delta_{ef}\,\{ade\}\,\{bce\}
\label{ortseij}
\end{equation}

\noindent where the notation $\{ade\}$ stands for the triangular delta
({\it viz.}, $\{ade\}$ is equal to $1$ if its three
arguments satisfy triangular inequalities, and is zero otherwise) and $\delta_{ef} \equiv \delta(e,f)$. \par 
\noindent The invariance under bistellar moves is related to the $PL$-equivalence class of the manifold involved.
Indeed, Pachner proved in \cite{pachner87} that two closed $n$-dimensional $PL$-manifolds are $PL$-homeomorphic
if, and only if, their underlying triangulations are related to each other by a finite sequence of bistellar moves. Thus,
in particular, the state sum (\ref{prsum1}) is an invariant of the  $PL$-structure of $M^3$.\par
\noindent Turning now to the case with boundary, we notice that (\ref{clsum2})  is manifestely invariant under 
bistellar moves which involve $3$-simplices in $int(T^3)$ (this is consistent with the remark that (\ref{clsum1})
reduces to (\ref{prsum1}) when $\partial M^3 =\emptyset$). In the non trivial case $\partial M^3 \neq  \emptyset$
new types of topological transformations have to be taken into account. Indeed
Pachner introduced moves which are suitable in the case of compact 
$n$-dimensional $PL$-manifolds with a non--empty boundary, the {\em elementary shellings} (see 
\cite{pachner90}). As the term "elementary shelling" suggests, this kind of operation 
involves the cancellation of one $n$-simplex (facet) at a time in a given triangulation $(T, \partial T) \rightarrow
(M, \partial M)$ of a compact $PL$-pair of dimension $n$. In order to be deleted, the facet must have some of its faces
lying in the boundary $\partial T$. It is possible to classify these moves according to the dimension of the components
of the facet in $\partial T$, and it turns out that there are just $n$ different types of elementary shellings in dimension $n$. 
Moreover, for each elementary shelling there exists an inverse move which corresponds to the attachment of a
new facet to a suitable component in $\partial T$.\par
\noindent In \cite{carbone} identities representing the three types of elementary shellings (and their inverse moves)
for a $3$-dimensional triangulation (\ref{coltri}) were established. 
Following the notation of \cite{russi} one of the identities is displayed below:\par 

\begin{eqnarray}
\lefteqn{\sum_{c\gamma} (2c+1)(-1)^{2c-\gamma}
\left( \begin{array}{ccc}
a & b & c \\
\alpha & \beta & \gamma
\end{array} \right)
\left( \begin{array}{ccc}
c & r & p \\
-\gamma & \rho ' & \psi
\end{array} \right)
 (-1)^{\Phi}\,
\left\{ \begin{array}{ccc}
a & b & c \\
r & p & q
\end{array}\right\}\,=}\hspace{1.5cm}\nonumber\\
& & =\,(-1)^{-2\rho}
\sum_{\kappa}\,(-1)^{-\kappa}
\left( \begin{array}{ccc}
p & a & q \\
\psi & \alpha & -\kappa
\end{array} \right)
\left( \begin{array}{ccc}
q & b & r \\
\kappa & \beta & -\rho '
\end{array} \right)
\label{clrodue}
\end{eqnarray}

\noindent where Latin letters $a,b,c,r,p,q,\ldots$ denote
angular momentum variables, Greek letters $\alpha, \beta, \gamma, \rho, \psi, \kappa,\ldots$ are the 
corresponding momentum projections and $\Phi \equiv a+b+c+r+p+q$.
The topological content of the above identity is the following: on the left--hand side there appears a tetrahedron,
two faces of which  (lying in $\partial T^3$)  share the edge labelled by $c$ and its projection $\gamma$; the
summation over $c$, $\gamma$ provide,
on the right--hand side, the appearance of the other two faces which survive after the shelling. The inverse move,
namely the attachment of a new facet to a pair of triangles in $\partial T^3$, is obtained by reading the
same identity backward.\par
\noindent Notice that the complete set of the identities can be actually derived
(up to suitable regularization factors) only from (\ref{clrodue}) 
and from both the orthogonality conditions for the $6j$'s (given in  (\ref{ortseij})) and
the completeness conditions for the $3jm$ symbols which read:\par

\begin{equation}
\sum_{c\gamma}\,(-1)^{2c -\gamma}\,(2c+1)
\left( \begin{array}{ccc}
a & b & c \\
-\alpha & -\beta & \gamma
\end{array} \right) 
\left( \begin{array}{ccc}
b & a & c \\
-\beta ' & -\alpha ' & \gamma
\end{array} \right)\,=\,
(-1)^{\alpha + \beta} \delta_{\alpha \alpha '}\: \delta_{\beta \beta '}
\label{ortrejm}
\end{equation}

\noindent The  recognition of the identities representing the elementary shellings and their inverse moves, 
together with a comparison
with the expression given in (\ref{clsum2}), allow us to conclude that 
the state sum $Z[(M^3, \partial M^3)]$  is formally
invariant both under (a finite number of) bistellar moves in the interior of $(M^3, \partial M^3)$ and under 
(a finite number of) elementary boundary operations, namely 
shellings and inverse shellings.  For what concerns $PL$- equivalence, we can exploit another result proved by Pachner in 
\cite{pachner90} which states that if $(T_1, \partial T_1) \rightarrow (M_1, \partial M_1)$ and 
$(T_2, \partial T_2) \rightarrow (M_2, \partial M_2)$ are
triangulations of $PL$, compact $n$-dimensional pairs, then
$|(T_1, \partial T_1)|\,\cong_{PL}\,|(T_2, \partial T_2)| \:\Longleftrightarrow\:(T_1, \partial T_1)\,\approx_{sh,bst}\,(T_2,
\partial T_2)$, where the equivalence $\approx_{sh,bst}$ is both under elementary shellings and under bistellar elementary
operations on $n$-simplices in $int(T_1)$ or $int(T_2)$. Thus (\ref{clsum1}) turns out to be an invariant of the
$PL$-structure (actually it is a topological invariant, since we are dealing with $3$-dimensional $PL$-manifolds). \par
\noindent As already stressed before, the complete set of elementary shellings can be derived from a single identity, 
namely (\ref{clrodue}), together with the conditions (\ref{ortseij}) and (\ref{ortrejm}) on the symbols.  
The structure of such identity strongly
rensembles the Biedenharn--Elliott identity (\ref{beidentity}), both for what concerns the number of symbols 
involved and owing to the presence
of a single sum over a $j$-variable. Recall also that the complete set of bistellar moves is actually derived from the B-E identity
+ (orthogonality conditions for the $6j$), apart from regularization. This  similarity in the algebraic structure of the
topological invariance in the two cases (closed and with boundary) is quite remarkable.\par
\noindent However, our previous analysis gives rise to other possible developments, in particular for what concerns
the search for state sum models of the same kind in dimension different from three. 
As we have already noticed, the different types of elementary shellings acting on a simplicial
$n$-dimensional pair $(T, \partial T)$ amount exactly to $n$, while the number of bistellar moves in the interior of 
$(T, \partial T)$ (or in $T$ itself if it is closed) is $(n+1)$. Moreover, the central projection of each elementary 
shelling onto $\partial T$ gives a particular bistellar move in dimension $(n-1)$ (being $\partial T$ a 
triangulation of a closed $(n-1)$-dimensional manifold). It is also easy to check that the same kind of projection 
of the complete set of boundary operations reproduces the complete set of bistellar moves in the lower dimensional case.
Starting from these remarks, we show in the following how a consistent $2$-dimensional state sum invariant arises
as {\em projection} of the invariant given in (\ref{clsum1}) and  (\ref{clsum2}).\par

\vskip 1 cm

Since the structure of a local arrangement of $2$-simplices 
in the state sum (\ref{clsum2}) is naturally encoded in (\ref{clrodue}), it turns out that a
state sum for a $2$-dimensional closed triangulation $T^2(j;m,m') \rightarrow M^2$ can be
defined by associating with each $2$-simplex $\sigma^2 \in T^2$ the following product of two $3jm$
symbols (a {\em double} $3jm$ symbol for short):\par

\begin{equation}
\sigma^2 \:\longleftrightarrow\: (-1)^{\sum_{s=1}^3 (m_s+m'_s)/2}
\left( \begin{array}{ccc}
j_1 & j_2 & j_3 \\
m_1 & m_2 & -m_3
\end{array}\right)
\left( \begin{array}{ccc}
j_1 & j_2 & j_3 \\
m'_1 & m'_2 & -m'_3
\end{array}\right)
\label{doublesim}
\end{equation}

\noindent where $\{m_s\}$ and $\{m'_s\}$ are two different sets of momentum projections associated with the
same angular momentum variables $\{j_s\}$, $-j \leq m_s, m'_s \leq j$ $\forall s=1,2,3$. The expression of the
state sum of the given triangulation reads:\par

\begin{eqnarray}
\lefteqn{Z[T^2(j;m,m')\rightarrow M^2; L] =}\hspace{.5in} \nonumber\\
& & =\Lambda(L)^{-N_0}\,\prod_{A=1}^{N_1} (2j_A+1) (-1)^{2j_A}(-1)^{-m_A-m'_A}\nonumber\\
& & \prod_{B=1}^{N_2} 
\left( \begin{array}{ccc}
j_1 & j_2 & j_3 \\
m_1 & m_2 & -m_3
\end{array}\right)_B
\left( \begin{array}{ccc}
j_1 & j_2 & j_3 \\
m'_1 & m'_2 & -m'_3
\end{array}\right)_B
\label{bisum1}
\end{eqnarray}

\noindent where $N_0, N_1, N_2$ are the numbers of vertices, edges and triangles in $T^2$, respectively.
Summing over all of the admissible assignements of $\{j;m,m'\}$ we get:\par

\begin{equation}
Z[M^2]\,=\,\lim_{L\rightarrow \infty}\:\sum_{\{T^2(j;m,m'), j \leq L\}}\:Z[T^2(j;m.m')\rightarrow M^2; L]
\label{bisum2}
\end{equation}

\noindent where the regularization is carried out according to the usual prescription.\par
\noindent In order to address the invariance of (\ref{bisum1}) we may start from a configuration representing 
two triangles glued together along a common edge $q$ with momentum projections $\kappa ,\kappa'$;
taking into account (\ref{clrodue}) and (\ref{ortseij}) we get\par

\begin{eqnarray}
\lefteqn{\sum_q \sum_{\kappa,\kappa'}
(2q+1)(-1)^{2q}\,(-1)^{-\kappa -\kappa'}
\left( \begin{array}{ccc}
p & a & q \\
\psi & \alpha & -\kappa
\end{array}\right)
\left( \begin{array}{ccc}
q & b & r \\
\kappa & \beta & \rho
\end{array}\right) \cdot}\nonumber\\
& & \cdot \left( \begin{array}{ccc}
p & a & q \\
\psi' & \alpha' & -\kappa'
\end{array}\right)
\left( \begin{array}{ccc}
q & b & r \\
\kappa' & \beta' & \rho'
\end{array}\right)
\,=\,\sum_c 
\sum_{\gamma, \gamma'} (2c+1)\,(-1)^{2c}\,(-1)^{-\gamma -\gamma'}\cdot\nonumber\\
& & \cdot \left( \begin{array}{ccc}
a & b & c \\
\alpha & \beta & \gamma
\end{array}\right)
\left( \begin{array}{ccc}
r & p & c \\
\rho & \psi & -\gamma
\end{array}\right)
\left( \begin{array}{ccc}
a & b & c \\
\alpha' & \beta' & \gamma'
\end{array}\right)
\left( \begin{array}{ccc}
r & p & c \\
\rho' & \psi' & -\gamma'
\end{array}\right)
\label{dueflip}
\end{eqnarray}

\noindent The geometrical meaning of the above identity should be clear: it represents the so called {\em flip} in dimension
$2$, namely the bistellar move $(2\rightarrow 2)$ involving a pair of triangles.\par
\noindent In order to obtain the remaining bistellar moves, we start from a configuration of three triangles glued along 
their edges $q,r,p$ in such a way that they share a common vertex. Using again (\ref{clrodue}), (\ref{ortseij}) and (\ref{ortrejm})
we find:\par

\begin{eqnarray}
\lefteqn{\sum_{q,r,p} 
(2q+1)(2r+1)(2p+1)\,(-1)^{2q+2r+2p}
\sum_{\kappa,\kappa'}\sum_{\rho,\rho'}\sum_{\psi,\psi'}
(-1)^{-\kappa -\kappa'}(-1)^{-\rho -\rho'} \cdot}\hspace{.5in}\nonumber\\
& & \cdot (-1)^{-\psi -\psi'} \left( \begin{array}{ccc}
p & a & q \\
\psi & \alpha & -\kappa
\end{array}\right)
\left( \begin{array}{ccc}
q & b & r \\
\kappa & \beta & -\rho
\end{array}\right)
\left( \begin{array}{ccc}
r & c & p \\
\rho & \gamma & -\psi
\end{array}\right)\cdot \nonumber\\
& & \cdot \left( \begin{array}{ccc}
p & a & q \\
\psi' & \alpha' & -\kappa'
\end{array}\right)
\left( \begin{array}{ccc}
q & b & r \\
\kappa' & \beta' & -\rho'
\end{array}\right)
\left( \begin{array}{ccc}
r & c & p \\
\rho' & \gamma' & -\psi'
\end{array}\right)\,=\nonumber\\
& & =\, \Lambda(L)^{-1}\,
\left( \begin{array}{ccc}
a & b & c \\
\alpha & \beta & \gamma
\end{array}\right)
\left( \begin{array}{ccc}
a & b & c \\
\alpha' & \beta' & \gamma'
\end{array}\right)
\label{duebary}
\end{eqnarray}

\noindent This identity represents the barycentric subdivision (and its inverse operation) of a triangle of edges $a,b,c$,
namely the bistellar moves denoted by $(1\leftrightarrow 3)$. As a matter of fact, the state sum given in (\ref{bisum1})
and (\ref{bisum2}) is formally invariant under (a finite number of) topological operations represented by (\ref{dueflip}) and
(\ref{duebary}). Thus, again from Pacner's theorem proved in \cite{pachner87}, we conclude that it is a $PL$-invariant 
(a topological invariant indeed).\par

\vskip 1 cm

The procedure oulined above for the $3$-dimensional model can be further generalized to the case of the quantum
enveloping algebra $U_{{\bf q}}(sl(2,{\Bbb C}))$, for ${\bf q}$ a root of unity. The corresponding {\em quantum}
invariant $Z_{{\bf q}}[M^3, \partial M^3]$ is the counterpart of the Turaev--Viro invariant  for
a closed $3$-dimensional $PL$-manifold $M^3$ (see \cite{turaev}) and details on its derivation can be found 
in \cite{carbone}. The same remark holds true for the $2$-dimensional closed model as well. Moreover, this extension
provide us with a {\em finite} topological invariant, which will  be easily evaluated and will turn out to be obviously related 
to the Euler characteristic of the closed suface.  \par 
\noindent  Without entering into technicalities, the quantum state sum which replaces (\ref{bisum1})
will contain the ${\bf q}$-analog of the double Wigner $3jm$ symbols 
(\ref{doublesim}) (with a suitable choice of normalization), while
the spin variables $j$'s take their values in a finite set $I=(0,1/2,1,\ldots,{\bf k})$ where $exp(\pi i/{\bf k})={\bf q}$.
For each $j \in I$ a function $w^2(j) \equiv w^2_j \doteq (-1)^{2x_j}[2x_j+1]_{{\bf q}} \in K^*$ is defined, 
where $K^* \equiv K \setminus \{0\}$ ($K$ a commutative ring with unity). Notice also that the notation
$[.]_{{\bf q}}$ stands for a ${\bf q}$-integer, namely $[n]_{{\bf q}}$  $= ({\bf q}^n-{\bf q}^{-n})/({\bf q}-{\bf q}^{-1})$
and that, for each admissible triple ($j,k,l$), we have: $w_j^{-2} \sum_{k,l}\,w^2_k w^2_l = w^2$, with $w^2=-2{\bf k}/
({\bf q}-{\bf q}^{-1})^2$. Coming back to the terms of the quantum state sum, we see that the classical 
weights $(-1)^{2j}(2j+1)$ in (\ref{bisum1}) are replaced by $w_j^2$, while each of the factors $\Lambda^{-1}$ becomes
$w^{-2}$. Then the state sum $Z_{{\bf q}}[T^2(j;m,m')\rightarrow M^2]$ for a given triangulation
can be easily evaluated, and it turns out that the associated quantum invariant amounts to:

\begin{equation}
Z_{{\bf q}}[M^2]\,=\,w^2\;w^{-2{\bf \chi}(M^2)}
\label{eulero}
\end{equation}

\noindent where ${\bf \chi}(M^2)$ is the Euler characteristic of the manifold $M^2$.\par
\noindent The above result shows that the $2$-dimensional closed model derived from our $3$-dimensional model
with a non empty boundary is not trivial, we actually recover the only topological invariant which is significant
for a closed surface.\par 
\noindent Following the same line of reasoning which gives rise to the closed $2$-dimensional model, 
we are currently addressing the analysis of the structure of the
{\em projected} counterpart of the algebraic identities representing topological moves in dimension four. 
We think that the approach oulined in this talk could provide us with a hierarchy of models (with and 
without boundary) from dimension four to dimension two.\par 

\vfill

\newpage

\section*{References}

\begin{description}

\bibitem[C-C-M]{carbone}
Carbone, G., Carfora, M., Marzuoli, A.:
Wigner symbols and combinatorial invariants of three--manifolds with boundary.
Preprint DFNT--T 14/98 and SISSA 118/98/FM

\bibitem[C-F-S] {carter95}
Carter, J.S., Flath, D.E., Saito, M.:
The Classical and Quantum {\it 6j}-symbols.
Math. Notes {\bf 43}. Princeton, NJ: Princeton University Press 1995

\bibitem[C-K-S] {carter98}
Carter, J.S., Kauffman, L.H., Saito, M.:
Structure and diagrammatics of four dimensional topological 
lattice field theories.
Preprint, math. GT/9806 023 (1998)

\bibitem[O92] {ooguri4d}
Ooguri, H.:
Topological lattice models in four dimensions.
Mod. Phys. Lett. A {\bf 7}, 2799-2810 (1992)

\bibitem[P87] {pachner87}
Pachner, U.:
Ein Henkeltheorem f\"{u}r geschlossene semilineare Mannigfaltig--
keiten.
Result. Math. {\bf 12}, 386-394 (1987)

\bibitem[P90] {pachner90}
Pachner, U.: 
Shellings of simplicial balls and p.l. manifolds with boundary.
Discr. Math. {\bf 81}, 37-47 (1990)

\bibitem[P-R] {ponzano}
Ponzano, G., Regge, T.:
Semiclassical limit of Racah coefficients.
In: Bloch, F. et al (eds.) Spectroscopic and Group Theoretical 
Methods in Physics, pp. 1-58.
Amsterdam: North-Holland 1968

\bibitem[R] {regge}
Regge, T.:
General Relativity without coordinates.
Nuovo Cimento {\bf 19}, 558-571 (1961)

\bibitem[R-S] {rourke}
Rourke, C., Sanderson, B.:
Introduction to Piecewise Linear Topology.
New York: Springer-Verlag 1982

\bibitem[T-V] {turaev}
Turaev, V., Viro, O.Ya.:
State sum invariants of 3-manifolds and quantum {\it 6j}-symbols.
Topology {\bf 31}, 865-902 (1992)

\bibitem[V-M-K] {russi}
Varshalovich, D.A., Moskalev, A.N., Khersonskii, V.K.:
Quantum Theory of Angular Momentum.
Singapore: World Scientific 1988

\bibitem[Y-L-V] {lituani}
Yutsis, A.P., Levinson, I.B., Vanagas, V.V.:
The Mathematical Apparatus of the Theory of Angular Momentum.
Jerusalem: Israel Program for Sci. Transl. Ltd. 1962

\end{description}

\end{document}